\documentclass[twocolumn,showpacs,preprintnumbers,amsmath,amssymb]{revtex4-1}

\usepackage{graphicx}
\usepackage{dcolumn}
\usepackage{bm}
\usepackage{bbm}

\newcommand{\be}{\begin{equation}}
\newcommand{\beq}{\begin{equation}}
\newcommand{\en}{\end{equation}}
\newcommand{\eeq}{\end{equation}}
\newcommand{\bea}{\begin{eqnarray}}
\newcommand{\ena}{\end{eqnarray}}
\newcommand{\hbo}{\hbox to 1 true cm {\hfill } }
\newcommand{\tr}{\hbox{tr}}

\begin{document}


\title{The density of states in gauge theories}

\author{Kurt Langfeld$^{a}$}
\author{Biagio Lucini$^b$}
\author{Antonio Rago$^a$}

\affiliation{%
\bigskip
$^a$School of Computing \& Mathematics,
Plymouth, PL4 8AA, UK  }

\affiliation{%
$^b$College of Science, Swansea University, Swansea,  SA2 8PP,  UK
}%

\date{ March 19, 2012
}

\begin{abstract}
The density of states is calculated for a SU(2) and a compact U(1)
lattice gauge theory using a modified version of the Wang-Landau
algorithm. We find that the density of states of the  SU(2) gauge theory can
be reliably calculated over a range of 120,000 orders of magnitude for lattice
sizes as big as $20^4$.  We demonstrate the potential of the algorithm by
reproducing  the SU(2) average action, its specific heat  and the critical
couplings of the weak first order transition in U(1). 
\end{abstract}

\maketitle

Monte-Carlo simulations~\cite{Creutz:1980zw} of the theory discretised
on a Euclidean space-time lattice~\cite{Wilson:1974sk} currently  provide the
most successful approach to calculations from first principles in asymptotically
free gauge theories in the energy domain in which the coupling is of
order one. Although this strategy is successful for computations
of observables that can be expressed as a vacuum expectation value
({\em vev}) on a theory with a semi-positive definite path integral measure,
when 
the observable is not a {\em vev} (e.g., the free energy, which is related
to the logarithm of a partition function) or the path-integral measure
is not semi-positive (like in QCD at finite density),
Monte-Carlo algorithms are either unsuitable or very inefficient.

An alternative numerical approach to Lattice Gauge Theories
potentially free from those limitations is based on the density of states.
Let us consider a quantum field theory with action $\beta S[\phi]$, with
$\beta$ the inverse 
coupling. For this theory, the path integral in Euclidean space-time is given by
\be
Z = \int {\cal D} \phi(x) \;  \mathrm{e}^{\beta S[\phi]} \ ,
\label{eq:pathintegral}
\en
where $\left( {\cal D} \phi(x) \right) $ means that the integral has
to be performed over all allowed configurations of the field $\phi$.
Defining the density of states $\rho(E)$ as 
\bea
\rho(E) &=& \int {\cal D} \phi(x) \;  \delta (S[\phi] - E) \ ,
\label{eq:2}
\ena
the path integral can be rewritten as
\bea
Z &=& \int \rho(E) \mathrm{e}^{\beta E} \, {\rm d} E \, ,
\label{eq:l} 
\ena
and the {\em vev} of an observable $O(E)$ becomes
\be
\langle O \rangle = \frac{1}{Z} \int \rho(E) \, O(E) \, \mathrm{e}^{ \beta E}
\, {\rm d} E \ .
\label{eq:3}
\en
If the density of states is known, the path integral and $\langle O
\rangle$ can be obtained by computing
numerically or analytically respectively the
integral~(\ref{eq:l})~and~(\ref{eq:3}). 

An efficient algorithm for computing $\rho(E)$ in systems with
discrete energy levels has been proposed by
Wang and Landau in~\cite{Wang:2001ab}.  To date, the method
has found various applications in Statistical Mechanics, some of which
have produced remarkable results that can not be obtained with a
direct Monte-Carlo approach (see e.g.~\cite{Hietanen:2011jy} for a recent
example). Despite its popularity in
Statistical Mechanics, the Wang-Landau algorithm has found only
limited applications in Lattice Gauge
Theory~\cite{Berg:2006hh,Bringoltz:2008zq}. In fact, the sampling of a
continuous density of
states with a straightforward generalisation of the method given
in~\cite{Wang:2001ab}  turns out to be
problematic~\cite{
Xu:2007aa,Sinha:2007aa}.
In this work, we propose a new method for determining a continuous density of
states and we apply it to calculate the density of states in SU(2) and U(1) on
the lattice.

Throughout this paper we adopt the lattice regularisation, which
leaves us with a $N^4$ cubic lattice as the discretisation of the Euclidean
space-time. The dynamical degrees of freedom of the SU$(N_c)$ gauge theory
are represented by the matrices $U_\mu (x) \, \in \,{\rm SU}(N_c)$, which are
associated with the links of the lattice. We are using the so-called
Wilson action, i.e.,
\be
S[U] = \sum _{\mu > \nu, x} \frac{1}{N_c} \, \mathrm{Re} \, \tr \Bigl[
U_\mu (x) \, U_\nu (x+\mu) \, U^\dagger _\mu (x +\nu ) \, U^\dagger _\nu (x)
\Bigr] ,
\label{eq:5}
\en
stressing however that our approach is not limited to this particular action, but can
handle e.g.~improved actions equally well.

%
\begin{figure}[t]
\includegraphics[height=7cm]{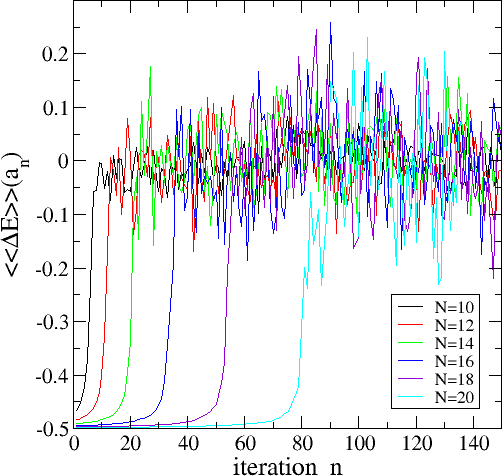} 
\caption{\label{fig:1} The thermalisation history for a SU(2) gauge theory
for lattice sizes $10^4 \ldots 20^4$.
}
\end{figure}
In order to present our novel type of numerical algorithm to
calculate the density of states, we will assume that $\ln \, \rho (E)$ is well
approximated by piecewise linear functions.
It will indeed turn out below that $\ln \, \rho (E)$ is a remarkable smooth function of $E$.

Let us consider the energy
interval $[E_0, E_0 + \delta E]$ for which we approximately write
\be
\rho (E) \; = \; \rho(E_0) \, \exp \Bigl\{ a(E_0) \, (E-E_0) \Bigr\}
\label{eq:6}
\en
for $E_0 \le E < E_0+ \delta E $.
Our goal will be to calculate the coefficients $a(E_0)$, which can be
considered as derivatives of the density of states:
\be
a(E_0) \; = \; \frac{{\rm d} \, \ln \rho (E) }{{\rm d}E}  \Big\vert _{E=E_0} \; .
\label{eq:7}
\en
The strategy to obtain these coefficients is based upon the truncated
and re-weighted expectation values defined by
\bea
{\langle \kern-.17em \langle} f(E) {\rangle \kern-.17em \rangle }(a)
&=& \frac{1}{\cal N} \int {\rm d}E \; f(E) \; \rho (E) \, \theta _{[E_0,\delta E]} \;
\; \mathrm{e}^{-aE} \; ,
\label{eq:9} \\
{\cal N} &=& \int {\rm d}E \, \rho (E) \, \theta _{ [E_0,\delta E]} \; \mathrm{e}^{-aE}
\; ,
\label{eq:10} \\
\theta _{[E_0,\delta E]} &=& \left\{ \begin{array}{ll}
1 & \hbox to 1 true cm {for \hfill }  E_0 \le E < E_0+\delta E , \\
0 & \hbox to 1 true cm {elsewhere.}  \end{array} \right.
\label{eq:11}
\ena
If the energy interval is small enough, i.e., if (\ref{eq:6}) is a good
approximation, we should be able to choose $a$ to compensate $a(E_0)$. This 
would leave us with a flat energy histogram and with
\be
{\langle \kern-.17em \langle} E {\rangle \kern-.17em \rangle }(a)
\; = \; E_0 + \frac{\delta E}{2} \; , \; \; \; \hbox{for} \; \; a = a(E_0).
\label{eq:15}
\en
Assume now that $a_n$ is an approximation for $a(E_0)$ such that
$x = [a(E_0)-a_n] \, \delta E \ll 1 $. Defining
$\Delta E := E - E_0 - \delta E /2 $, we then find using (\ref{eq:6})
\be
{\langle \kern-.17em \langle} \Delta E {\rangle \kern-.17em \rangle }(a_n)
\; = \; \frac{\delta E^2}{12} \, [a(E_0) - a_n] \; + \;
{\cal O }(x^3 \, \delta E) \; .
\label{eq:17}
\en
\begin{figure}[t]
\includegraphics[height=7cm]{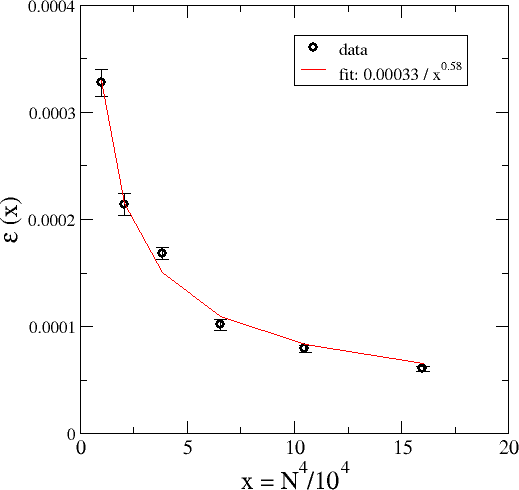} 
\caption{\label{fig:2} The statistical error for the estimate of
$a(E_0)$ for lattice sizes $10^4 \ldots 20^4$.
}
\end{figure}
Ignoring the higher order correction and solving for $a(E_0)$, we
obtain a better approximation $a_{n+1}$:
\be
a_{n+1} \; = \; a_n \; + \; \frac{12}{\delta E ^2} \,
{\langle \kern-.17em \langle} \Delta E {\rangle \kern-.17em \rangle }(a_n)
\; .
\label{eq:18}
\en
The central idea is to iterate the latter equation until
$$
{\langle \kern-.17em \langle} \Delta E {\rangle \kern-.17em \rangle }(a_\infty )
\; = \; 0 \; \; \Rightarrow \; \; a_\infty = a(E_0) \; ,
$$
where we have used (\ref{eq:15}). We point out that the truncated expectation
values can be easily estimated by means of Monte-Carlo methods.
To this aim, we insert (\ref{eq:2}) into (\ref{eq:9}) to obtain:
\bea
{\langle \kern-.17em \langle} f(E) {\rangle \kern-.17em \rangle }(a)
&=& \frac{1}{\cal N} \int_{[E_0,\delta E]} {\cal D}U_\mu \;
f\Bigl(S[U] \Bigr) \; \mathrm{e}^{-a S[U]} \; ,
\label{eq:19} \\
{\cal N} &=& \int_{[E_0,\delta E]} {\cal D}U_\mu \; \mathrm{e}^{-a S[U]} \; .
\label{eq:20}
\ena
The subscript of the integral indicates that updates of configurations
the action of which falls outside the desired energy interval are
discarded. There are many Monte-Carlo techniques to estimate the
truncated expectation value in (\ref{eq:19}), the Metropolis algorithm and
the Heat-Bath approach being the two most obvious choices. We have tested both 
techniques and found that our method for estimating $a(E_0)$ is robust. The
numerical results shown below have been obtained by an adapted Heat-Bath
algorithm with a 100\% acceptance rate (details of the algorithm will be
published in a forthcoming paper).

%
\begin{figure}[t]
\includegraphics[height=7cm]{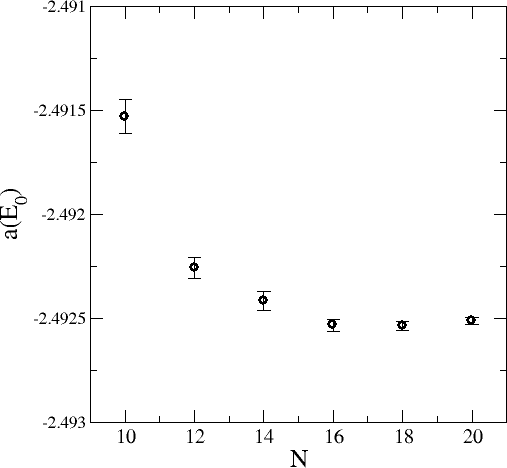} 
\caption{\label{fig:3} The estimates for $a(E_0)$ for $E_0 = 0.650 \, \times
  \,  6N^4$
as a function of the lattice size.
}
\end{figure}
\begin{figure}
\includegraphics[height=7cm]{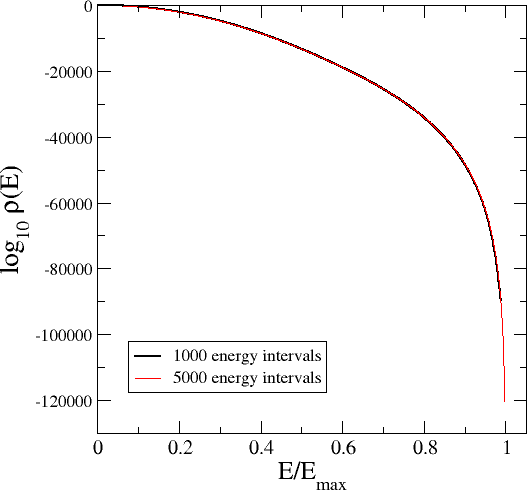} 
\caption{\label{fig:4} The logarithm (base 10) of the density of states for the
SU(2) gauge theory using a $10^4$ lattice.
}
\end{figure}
\begin{figure}
\includegraphics[height=7cm]{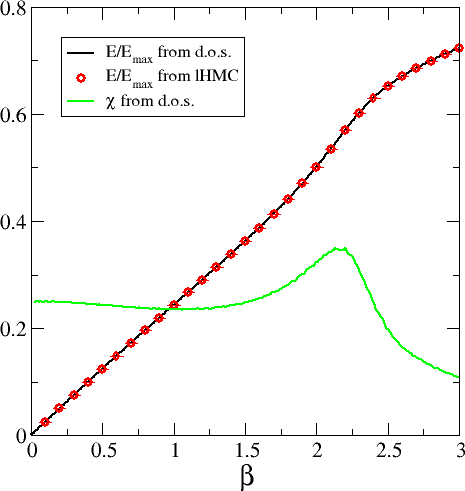} 
\caption{\label{fig:5} Average plaquette for a SU(2) gauge theory on a
$10^4$ lattice obtained by means of the density of states and
local-hybrid Monte-Carlo. Also shown is the specific heat $\chi (\beta )$,
Eq.~(\ref{eq:26}).
}
\end{figure}
Let us now consider the SU(2) gauge theory to illustrate our approach in
practice. If $N^4$ is the number of lattice points, the maximal action
is given by $E_\mathrm{max} = 6 N^4$. We here consider the energy interval
$I:=[E_0, E_0+\delta E] = [0.650,0.651] \, 6N^4$. The first task is to
generate a lattice configuration $\{U_\mu \}$ the action of which falls
into the energy interval $I$. For this purpose, we start with a ``cold''
configuration $U_\mu(x)=1$, and update the configuration forcing it to
reach the desired energy interval. We then pick a start value for the iteration (\ref{eq:18}), which has been $a_0=-2$ in this preliminary study.
We perform $25$ energy restricted
Monte-Carlo sweeps at $a_0$ (see (\ref{eq:19})), where each sweep consists of
$N^4$ updates of randomly chosen individual links.

%
In order to evaluate the next $a_i$  the expectation value
${\langle \kern-.17em \langle} \Delta E {\rangle \kern-.17em \rangle
}$ is evaluated using
the energy restricted Monte-Carlo method  (see (\ref{eq:19})). For this, we have used 
$384$ measurements divided in $48$ independent runs each
contributing $8$ Monte-Carlo sweeps (these calculation are performed on
the HPC computing facilities at the Plymouth University). The corresponding
estimator is then used to obtain an improved value $a_1$. 
This procedure is reiterated $n$ times, $n > 1$, until the value of $a$ starts to fluctuate around a central value. The thermalisation history is shown
in figure~\ref{fig:1}: for small lattice sizes such as $10^4$, a thermalised state
is reached after $10$ iterations while for our biggest lattice $20^4$
roughly $80$ iterations are necessary to reach an equilibrium.
To keep control of the autocorrelation in the determination of the
solution of the
iterative procedure we have evaluated the integrated autocorrelation
time ($\tau_{int}$) of ${\langle \kern-.17em \langle} \Delta E {\rangle \kern-.17em \rangle }$.
In particular the measure of $\tau_{int}$ for the highest energy gap
yields a value always smaller that two steps for each of our volumes.

Having control of the autocorrelation time allows us to reliably
define a statistical error of
${\langle \kern-.17em \langle} \Delta E {\rangle \kern-.17em \rangle }$
which directly feeds into the uncertainty for $a_{n+1}$ (see (\ref{eq:18})).
Rather than to spend all numerical resources to obtain a high-precision
estimate for
${\langle \kern-.17em \langle} \Delta E {\rangle \kern-.17em \rangle }$
we found is advantageous to feed the more noisy estimator into the
iteration (\ref{eq:18}) and to average the $a_n$ values of the
resulting sequence. The standard error of $a_n$ for an
average over a bin of $10$ iterations {\it after} thermalisation is
shown in figure~\ref{fig:2}. We roughly find that the error decreases
like $1/\sqrt{V}$ where $V$ is the lattice volume. The lack of
autocorrelation reflects in the  good scaling of the error with the
volume showing the efficiency of the algorithm also for large
volumes. In particular, this observation is true even when studying energy intervals for which we would normally
expect strong effects in autocorrelation due to critical slowing down
(for example for  $ 0.850 \leq   E/E_{\mbox{max}}  \leq 0.851$ and $V=20^4$ we
find $\tau_{int}=1.8(1)$).

For the determination of $a(E_0)$, $187$ iterations have been
performed for thermalisation and $312$ further iterations were carried out
to estimate $a(E_0)$. Our findings as a function of the lattice size
are shown in figure~\ref{fig:3}.

%
\begin{figure}[h]
\includegraphics[width=7.4cm]{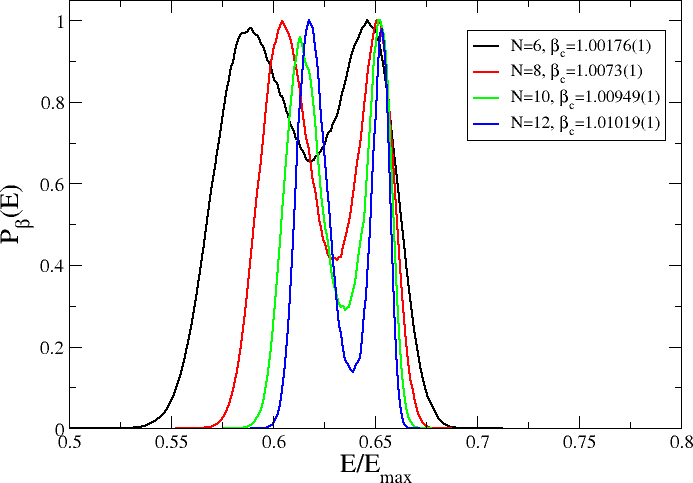} 
\caption{\label{fig:7} The probability density $P_\beta (E)$ for a compact
  $U(1)$ gauge theory at critical coupling for several lattice sizes
  $N^4$.
}
\end{figure}

Once $a(E_0)$ has been obtained  for all energies $E_0^i = i \times
\delta E $ (we here only consider positive
  energies), the density of states $\rho (E)$ can be easily constructed
from (\ref{eq:6}):
\be
\rho (E) = \prod _{i=1}^k \mathrm{e}^{a(E_0^i) \, \delta E } \,
\exp \Bigl\{ a(E_0) \, (E-E_0^k) \Bigr\}
\label{eq:25}
\en
for $E_0^k \le E < E_0^{k+1}$. Thereby, we have normalised the density
of states such that $\rho (E=0)=1$. Our numerical result is shown in
figure~\ref{fig:4}. In order to estimate any influence of the discretisation
error, we have calculated the density of states by splitting the
energy interval $[0,E_\mathrm{max}]$ into $1000$ and $5000$ energy
intervals. Both curves fall on top of each other in figure~\ref{fig:4}.
As a proof of concept that our numerical approach does yield
high precision expectation values, we have calculated the average
plaquette $\langle E \rangle / E_\mathrm{max}$ using (\ref{eq:3}).
As expected, only a small energy window with $a(E) \approx \beta $
significantly contributes to the expectation value. Care has been taken
to handle potentially large numbers. We have compared our result with
that from a standard method using local-hybrid Monte-Carlo. A very good
agreement is observed. An observable which is generically difficult to
estimate due to cancellations is the specific heat, which we define by
\be
\chi (\beta ) \; = \; \frac{1}{6N^4} \, \Bigl(
\langle E^2 \rangle \, - \, \langle E \rangle ^2 \Bigr) ,
\label{eq:26}
\en
where the expectation values are obtained by means of (\ref{eq:3}).
Our numerical findings for $\chi $ are also shown in figure~\ref{fig:5}.
We have checked for a few $\beta $ values that our result agrees with
that obtained by standard methods.

We have finally tested our approach for the compact U(1) gauge with Wilson
action (\ref{eq:5}). Here, the links are U(1) group elements, i.e.,
$U_\mu (x) = \exp \{ i \theta _\mu (x) \} $ with $\theta _\mu (x) = -\pi \ldots
\pi $ being the dynamical degrees of freedom featuring in the functional
integral with a constant measure. By means of a large scale investigation on the
basis of the Borgs-Kotecky finite size scaling analysis, it has been
finally established in~\cite{Arnold:2002jk} that compact U(1) possesses a weak
first-order phase  transition at $\beta = \beta _c \approx 1.0111331(21) $ (in
the infinite volume limit).  An unmistakable sign for a first order transition
is the characteristic double-peak structure in the action probability density,
i.e.,
\be
P_\beta (E) \; = \; \rho (E) \, \exp \{ \beta E \} \; ,
\label{eq:30}
\en
for $\beta \to \beta _c$. It turns out that this double-peak
structure is very sensitive to variations of $\beta $ allowing a high
precision determination of $\beta _c$ at finite volume, i.e., the critical 
coupling for which the peaks are of equal height. Note that we have normalised
$P_\beta (E)$ such that its maximum value equals one. The critical couplings
$\beta _c$, listed in the graph, are in good agreement with those
from the large scale study~\cite{Arnold:2002jk}.

In conclusions, we have developed a modified version of the Wang-Landau
algorithm suitable for theories with continuous degrees of freedom.
We have shown that the density of states for a SU(2) gauge theory
can be calculated over a range of 120,000 orders of magnitudes even
for a lattice as large as $20^4$. Our approach reliably reproduces the critical
couplings of the weak first order transition of the compact U(1) gauge theory.
Using the Cabibbo Marinari method~\cite{Cabibbo:1982zn}, our approach can
be generalised to SU$(N_c)$ Yang-Mills theories. Quantities of interests
which are earmarked for our approach are thermodynamic
potentials~\cite{Giusti:2011kt}, vortex free energies~\cite{Kovacs:2000sy}
and electric fluxes for the study of the mass-gap and
confinement~\cite{deForcrand:2001nd}. Finally, we point out that the statistical error for
expectation values obtained by the density of states method can be
obtained by the bootstrap technique. A careful investigation of
the statistical and possible systematic errors (from which our
results seem to be free) will be reported elsewhere.

\noindent {\bf Acknowledgments:}
This work is supported by STFC under the DiRAC framework. We are
grateful for the support from the HPCC Plymouth, where the numerical
computations have been carried out. BL is supported by the Royal
Society and by STFC.

\bibliography{density_ym}{}

\end{document}